\documentclass[twocolumn,amsmath,amssymb,prl]{revtex4}

\usepackage{graphicx} % Include figure files
\usepackage{dcolumn}  % Align table columns on decimal point
\usepackage{bm}       % bold math

\begin{document}

\draft

\title{ 
Fermiology of Cuprates from First Principles:\\
From Small Pockets to the Luttinger Fermi surface
}

\author{L. Hozoi, M. S. Laad and P. Fulde}
\address{Max-Planck-Institut f\"{u}r Physik komplexer Systeme,
         N\"{o}thnitzer Str.~38, 01187 Dresden, Germany}
\date\today


\maketitle

%% \begin{abstract}

\textsl{
Fermiology, the shape and size of the Fermi surface, underpins the low-temperature
physical properties of a metal.
Recent investigations of the Fermi surface of high-$T_{c}$ superconductors, however,
show a most unusual behavior:
upon addition of carriers, ``Fermi'' pockets appear around nodal (hole doping)
and antinodal (electron doping) regions of the Brillouin zone in the ``pseudogap''
state.
With progressive doping, $\delta$, these evolve into well-defined Fermi surfaces 
around optimal doping ($\delta_{opt}$), with no pseudogap.
Correspondingly, various physical responses, including $d$-wave superconductivity,
evolve from highly anomalous, up to $\delta_{opt}$, to more conventional beyond.
Describing this evolution holds the key to understanding high-temperature
superconductivity.  
Here, we present {\it ab initio} quantum chemical results for cuprates,
providing a quantitative description of the evolution of the Fermi surface
with $\delta$.
Our results constitute an {\it ab initio} justification for several, hitherto
proposed semiphenomenological theories, offering an unified basis for understanding
of various, unusual physical responses of doped cuprates.
}
 
%% \end{abstract}

%% \maketitle

\  

 Understanding high-temperature superconductivity in quasi two-dimensional
(2D), doped copper oxides remains one of the most challenging problems in condensed
matter physics. 
In spite of varying structural and chemical details, the phase diagram of the high-$T_{c}$
superconductors (HTS's) is seemingly remarkably universal: the undoped 
compounds with nominally one hole per Cu site are Mott insulators (MI's) due to strong 
electron-electron interactions \cite{CuO_PALee_06}.
Upon addition of charge carriers (doping), the cuprates turn into $d$-wave
superconductors ($d$-SC) at low temperatures, $T\!<\!T_c$
\cite{CuO_PALee_06}.

The ``normal'' state for $T\!>\!T_{c}$ is actually very abnormal, and radically
undermines the conventional Landau theory of Fermi liquids (FL's).
In the so-called underdoped (UD) regime, $\delta\!\ll\!1$, a $d$-wave pseudogap 
($d$-PG) characterises the normal state \cite{CuO_PALee_06}.
Whether this $d$-PG state is the precursor of $d$-SC at lower $T$ or its competitor 
is hotly debated \cite{CuO_NE_Ong_06,CuO_PG_varma06}.
Around optimal doping, a ``strange metal'' phase, with most unusual singular
responses \cite{CuO_QC_vdMarel_03} is clearly revealed: this is the celebrated non-FL metallic
state that has been investigated for twenty years \cite{CuO_PALee_06}.
In the overdoped (OD) regime ($\delta\!>\!\delta_{opt}$), low-$T$ FL behavior seems to 
be smoothly recovered.

Very recently, notable improvements in sample quality as well as measuring technology
have finally allowed accurate mapping of the actual dispersion of the quasiparticles
(QP's) and the Fermi surface (FS) of HTS's. 
Specifically, angle-resolved photoemission (ARPES)
\cite{CuO_damascelli_rev,CuO_tanaka06,CuO_kanigel06} and quantum oscillation (SdH)
techniques \cite{CuO_HP_doiron07,CuO_HP_yelland07} reveal crucial, hitherto unmapped features
of the evolution of the ({\it renormalised}) QP dispersion as a function of doping.
Thus, these works open up the possibility, for the first time, of unearthing the
link between the electronic structure and physical responses of cuprates in microscopic
detail as a function of $\delta$. 
Both ARPES and SdH measurements reveal a full FS
consistent with conventional band-structure calculations for $\delta\!>\!\delta_{opt}$
\cite{CuO_damascelli_rev,CuO_OD_hussey06}. However, in the UD regime, the 
small ``Fermi'' pockets inferred by SdH experiments are in deep conflict with
Luttinger's theorem.
For hole doped samples with $\delta\!=\!0.1$, the SdH results yield a carrier concentration
$x_{\rm SdH}\!=\!0.15$ \cite{CuO_HP_doiron07}.
Under the same conditions, the low-$T$ Hall constant is {\it electron}-like
\cite{CuO_HP_leboeuf07}.
How can this come about\,?  Existing theoretical calculations cannot resolve this
issue satisfactorily. And yet, this finding points toward a glaring discrepancy in our 
understanding of the electronic structure of cuprates.
In light of these findings, a consistent theoretical scenario aiming to describe the
unique physics of HTS's {\it must} now base itself upon the appropriate, collective
excitations stemming from the observed, detailed shape and size of the FS.

Here, we study the dispersion of the lowest hole and electron-addition states,
as well as the evolution of the {\it renormalised} FS with doping.
Using an {\it ab initio} wavefunction-based formalism, we describe these with
quantitative accuracy vis-a-vis recent ARPES and SdH measurements. 
Implications of our findings for other experiments, as well as their connection
to earlier model-based and semiphenomenological theories, are discussed in
detail.
Our findings lend credence to the view \cite{CuO_RVB_97}, that the unique
properties of cuprates are those of a 2D, doped MI.

\section{Theoretical framework}

The correlation-induced renormalisation effects on the valence and conduction energy
bands are remarkably strong in cuprates.
Early attempts to describe these effects were based on the $t$-$J$ model
\cite{CuO_PALee_06} and indicated the crucial role played by the strong antiferromagnetic
(AF) couplings in reducing the effective bandwidths.
In the three-band context, it was suggested that a doped oxygen hole would induce short-range ferromagnetic 
(FM) correlations between adjacent Cu sites \cite{FMpolaron_emery88}.
If the extra hole delocalizes over all four equivalent ligands of a given CuO$_4$ plaquette
\cite{ZR_88}, these FM correlations would involve Cu sites on five plaquettes 
\cite{CuO_hozoi_07}, as shown in Fig.~1.
Since the mobility of this entity is expected to be small, it is often referred to
as a FM ``spin polaron''.

%% FIGURE 1
\begin{figure}[b!]
\includegraphics*[angle=270,width=1.00\columnwidth]{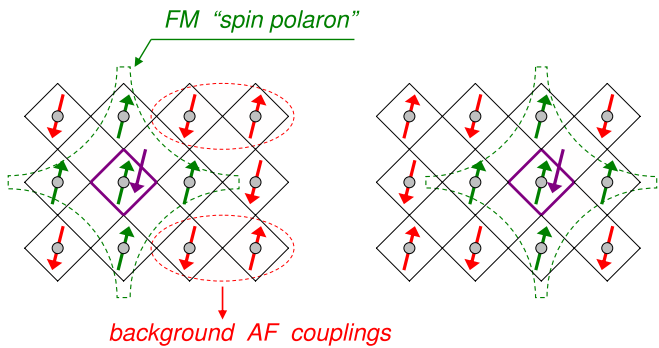}
\caption{Ferromagnetic correlations (green) among Cu sites around a doped oxygen
hole (violet) as inferred from CASSCF calculations.
Farther spin couplings are antiferromagnetic (red).
The motion of the oxygen hole is coherent when the FM spin
polarization ``cloud'' at adjacent Cu sites moves solidarily with the hole.
The O ligands are at the vertices of the square plaquettes, the Cu ions are
shown as grey dots.}
\end{figure}

An accurate investigation of the structure of such composite
objects calls for methods that allow an unbiased treatment of the various
(competing) interactions in the CuO$_2$ plane.
A fundamental point underlying the physics of cuprates is the interplay 
between electron localization effects as a result of strong repulsive interactions
and band-like behaviour as a result of translational symmetry and inter-site
orbital overlap.
Standard band theories based upon the density-functional model (and the local density
approximation, LDA) mainly emphasize the latter aspect.
Though LDA provides rather good results for weakly correlated solids, its limited ability
to describe correlated $d$ (and $f$) electrons is well-documented.
With the advent of dynamical mean-field theory (DMFT), this basic conflict has been
partially resolved \cite{dmft_rev_GK06}.
In particular, much progress in describing the FS's of real materials, along with 
their one-particle spectral functions, has been possible.
However, this is still some distance from being a totally {\it ab initio} approach,
since the actual correlations are approximated by local (Hubbard) {\it parameters}.
Use of constrained LDA to estimate these parameters entails an uncertainty of the
order of $20\%$, while their self-consistent estimation within LDA+DMFT is fraught
with insurmountable problems \cite{dmft_U_imada05}.

An alternative approach bases itself on state-of-the-art quantum chemical (QC) methods
\cite{fulde_wf_02}.
In molecular systems, wavefunction-based quantum chemistry 
provides a rigorous theoretical framework for addressing the electron correlation
problem \cite{QC_book_00}.
A real-space, QC-based treatment is then a natural starting point in dealing with
Mott physics in $d$-metal solid state compounds. 
As shown below, the $\mathbf{k}$-dependent energy bands can be recovered at a later stage
after rigorously accounting for the ubiquitous strong short-range correlation effects.

The strategy is to use a sufficiently large cluster, $\mathcal{C}$, cut out from the
infinite solid and properly embedded in some effective lattice potential, capable of 
describing these crucial short-range correlations accurately.
The presence of partially filled $d$ electron shells requires a multiconfiguration 
representation of the many-electron wavefunction.
The complete-active-space (CAS) self-consistent-field (SCF), CASSCF, method 
\cite{casscf_roos80} provides precisely such a framework (see Methods for details).
It can, for example, describe spin correlation effects such as the Anderson superexchange
in MI's \cite{thesis_Coen} and the double-exchange in mixed-valence systems
\cite{DE_alex06}.
For undoped cuprates, with formally one $3d_{x^{2}-y^{2}}$ electron per Cu site,
the CAS wavefunction is similar to the variational wavefunction used in numerical
studies of the 2D, one-band Hubbard model \cite{HubbM_sorella_05}.
However, {\it all} one- and two-particle integrals are computed here in a totally {\it ab initio} way.

The dispersion of $d$-like states on a square lattice is given by the
following relation:
\begin{eqnarray*}
\epsilon(\mathbf{k}) = &-&2t\,(\cos k_xa  + \cos k_ya) + 4t'\cos k_xa \cos k_ya \\
                       &-&2t''(\cos 2k_xa + \cos 2k_ya) \,,
\end{eqnarray*}
where $t$, $t'$, $t''$ are the hopping integrals between nearest-neighbor (NN),
second-NN and third-NN sites and the effective site is one CuO$_4$ plaquette
\cite{ZR_88}.
LDA calculations yield $t$ values of 0.4--0.5 eV and a ratio
between the NN and second-NN hoppings $t'/t\!\approx\!0.15$ for La$_2$CuO$_4$
and 0.33 for Tl$_2$Ba$_2$CuO$_6$ \cite{CuO_oka_01}.
In contrast, the CASSCF calculations predict a {\it renormalised} NN hopping $t\!=\!0.135$
eV in La$_2$CuO$_4$ \cite{CuO_hozoi_07}.
Here, we describe how the detailed QP dispersion can be obtained with quantitative
accuracy, for {\it both} hole and electron-addition states.
The effective hoppings are computed by using the overlap, $S_{ij}$,
and Hamiltonian, $H_{ij}$, matrix elements between $(N\!\mp\!1)$-particle
wavefunctions having the additional particle (hole or electron) located on different
plaquettes ($i$,$j$,...) of a given cluster.
Each of these $(N\!\mp\!1)$ wavefunctions, $|\Psi_{i}^{N\mp 1}\rangle$, is obtained
by {\it separate} CASSCF optimizations.
A similar scheme was previously applied to simpler, noncontroversial systems such
as diamond, silicon and MgO \cite{CBs_uwe_06,CBs_MgO_07}.
It accounts for both charge \cite{CBs_uwe_06,CBs_MgO_07} {\it and} spin
polarization and relaxation effects in the
nearby surroundings, see Fig.~1. 
For degenerate (i.e., $H_{ii}\!=\!H_{jj}$) $(N\!\mp\!1)$ states,
$t = (\epsilon _j - \epsilon _i)/2 = (H_{ij} - S_{ij} H_{ii}) / (1 - S_{ij}^2)$,
where $\epsilon _i$ and $\epsilon _j$ are the eigenvalues of the $2\!\times\!2$
secular problem.
For non-degenerate ($H_{ii}\!\neq\!H_{jj}$) states,
$t = 1/2[(\epsilon_j - \epsilon_i)^2 - (H_{jj} - H_{ii})^2]^{1/2}$.
The $S_{ij}$ and $H_{ij}$ terms are computed using the State-Interaction (SI)
method \cite{SI_Malmqvist_86}.
Correlation effects beyond CASSCF on the onsite matrix elements $H_{ii}$ are
calculated by multiconfigurational second-order perturbation theory, CASPT2
\cite{CASPT2_90}.
The short-range magnetic correlations are included in our clusters by adding extra
CuO$_4$ units around those plaquettes directly involved in the hopping process,
see Fig.~2.

%% FIGURE 2
\begin{figure}[b]
\includegraphics*[angle=270,width=0.98\columnwidth]{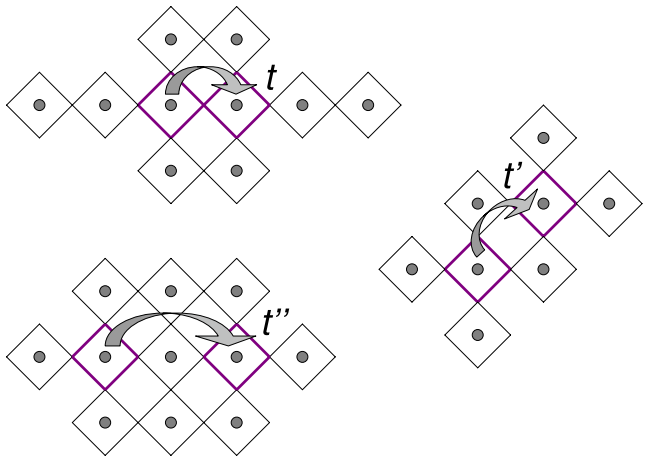}
\caption{Sketch of the finite clusters employed for the calculation of the effective
valence-band and conduction-band hoppings.}
\end{figure}

\section{Quasiparticle bands and ARPES}

We investigated both, the $p$-type HTS
La$_2$CuO$_4$ and the $n$-type HTS SrCuO$_2$.
Renormalised hopping matrix elements (ME's) for Zhang-Rice (ZR) \cite{ZR_88} type states
in La$_2$CuO$_4$ and electron-addition states in SrCuO$_2$, involving neighbors up to 
the third order, are listed in Table I.
For comparison, ``unrenormalised'' (or bare) hoppings were also computed, by imposing
a FM arrangement of spins at the nearby Cu sites, i.\,e., a FM lattice.
The bare hoppings are substantially smaller for the $(N\!+\!1)$ 
Cu $d^{10}$ states because the Cu $3d$ functions are more compact as compared to the O
$2p$ orbitals.

In effective one-band models \cite{ZR_88,CuO_PALee_06}, the ZR $p$-$d$ state is 
regarded as a vacant, or unoccupied, $d$-like site.
Consequently, there is no renormalisation of the second-NN and third-NN hoppings $t'$
and $t''$ because these connect sites of the same magnetic sublattice.
In contrast, the interplay between short-range FM correlations and longer-range AF
couplings (see the sketch in Fig.~1) produces large renormalisation effects for
{\it all} hopping ME's in our approach.
For $t'$ and $t''$, in particular, the hopping of the $2p$ hole implies coupled, Cu
and O spin ``flips'' on the two plaquettes directly involved in the hopping process.
Nevertheless, spin correlations decay rapidly with distance and so the renormalisation
effects are less drastic for the third-NN ME, $t''$.
We thus find $t''\!\simeq\!t/2$ and $t'\!\ll\!t$, a rather
remarkable result.

For the description of the electron-addition $d^{10}$ states, an effective one-band
model is seen to be justified.
As shown in Table I, in this case only $t$ is substantially affected by
nonlocal spin correlations.
Hence, for the ($N\!+\!1$) states, the renormalized hoppings satisfy $t'\!\simeq\! t$,
an equally remarkable result.
The particle-hole asymmetry, see Fig.~3, is now readily understood from the very different 
$t'/t$ and $t''/t$ values for the $(N\!-\!1)$ (ZR-like) and 
$(N\!+\!1)$ (Cu $d^{10}$) bands.

Doping of the CuO$_{2}$ planes is achieved by chemical substitution in the
``reservoir'' layers.
The dopant carriers must quantum mechanically tunnel from the reservoir to the
planes:
this necessarily involves the apical O $2p_z$, Cu $3d_{z^{2}}$ and Cu $4s$
orbitals, causing additional renormalisation of the planar QP's from these apical
charge-transfer interactions. 
Hence, we extended our calculations to include configurations where an electron
is removed/added from/to the Cu $3d_{z^{2}}$ or Cu $4s$ orbital.  
Only the Cu $3d_{z^{2}}$ $(N\!-\!1)$ state gave rise to considerable renormalisation
of the planar QP dispersion.
While the onsite mixing between the ZR and $d_{z^2}$ $(N\!-\!1)$ configurations and
NN hopping between degenerate $d_{z^2}$ hole states are negligible, the {\it inter-site}
off-diagonal hopping is large.
With sets of orbitals individually optimized for each $(N\!-\!1)$ state, this
off-diagonal ME is $t_m\!=\!0.20$ eV, larger than the value reported in
ref.~\cite{CuO_LH_MSL_07}, where the lowest $d_{z^2}$ hole state was expressed in 
terms of orbitals optimized over an average of several excited states involving 
different couplings among the nearby Cu spins. 
In $\mathbf{k}$-space, the hybridisation ME between the ZR and $d_{z^2}$ bands reads
$\gamma_{m}(\mathbf{k})\!=\!t_m (\cos k_x a\!-\!\cos k_y a)$.
Further, onsite, the ZR and $d_{z^2}$ $(N\!-\!1)$ states are separated by an energy
$\Delta\epsilon$.
It turns out that the correlation-induced corrections to the CASSCF energy separation
are substantial, changing this quantity from 0.60 eV \cite{CuO_LH_MSL_07} to
$\Delta\epsilon\!=\! 1.70$ at the CASPT2 level.
Such corrections are usually small, for
both the onsite {\it relative} energies \cite{thesis_Coen} and hoppings
\cite{CuO_hozoi_07,CBs_MgO_07}.
Nevertheless, corrections as large as 0.9 eV have been found before for the relative
energy of the $^{1}\!A_{1g}$ state of the $d^8$ manifold in NiO \cite{thesis_Coen},
for example.
 
It is now straightforward to diagonalise the $\mathbf{k}$-dependent $2\!\times\!2$
matrix,
\[ \left( \begin{array}{cc}         
          \epsilon_{{\rm ZR}}(\mathbf{k})  &\gamma_{ m} (\mathbf{k})  \\
          \gamma_{m}(\mathbf{k})           &\epsilon_{z^2}(\mathbf{k}) + \Delta\epsilon
          \end{array}
   \right)\,,     
\]
to yield the renormalised bands.          
This constitutes a non-trivial extension of the three-band Hubbard model,
where the additional renormalisation from the apical link is not considered.
The resulting dispersion of the ZR-like band is plotted in Fig.~3 and shows
excellent agreement with the dispersion of the lowest ARPES band reported for
La$_{2}$CuO$_{4}$ by Ino {\it et al.}~\cite{CuO_ARPES_ino00}.
In particular, the flat dispersion around (0,$\pi$), the maximum near ($\pi/2$,$\pi/2$)
and a renormalized bandwidth of nearly 1 eV are all faithfully reproduced in the
theoretical results.
We have not attempted to describe the ``waterfall''-like structures observed recently
in ARPES.
First, their interpretation \cite{CuO_WF_fink07} and causal link to $d$-SC are 
controversial.
Theoretically, the study of such structures requires an analysis of the {\it incoherent}
part of the spectral function \cite{dmft_rev_GK06,kakehashi_06}.
This challenging exercise is beyond the scope of our present work.
For the Cu $d^{10}$ states, a lack of detailed data for ARPES lineshapes in $n$-type 
cuprates precludes a direct comparison between theory and experiment.

%% TABLE 1
\begin{table}[t]
\caption{
Hopping ME's for ZR-like states in La$_2$CuO$_4$ and
electron-addition Cu $d^{10}$ states in SrCuO$_2$.
The bare hoppings were computed by imposing high-spin couplings among
the nearby Cu sites.
For the $d^{10}$ states, each of the hoppings changes by
less than $15\%$ from La$_2$CuO$_4$ to SrCuO$_2$ (not shown in the table).
On the other hand, no ZR-like solution was obtained for the $(N\!-\!1)$
states in SrCuO$_2$, which qualitatively confirms the experimental findings:
for in-plane lattice constants $a\!\ge\!3.87$ \AA \ ($a\!=\!3.925$ \AA \ in SrCuO$_2$),
the CuO$_2$ planes do not readily accept holes in the O bands \cite{SrCuO_JBG_91}.
}
\begin{ruledtabular}
\begin{tabular}{lcc}
Hopping ME's      &Bare             &Renormalised \\

\colrule
\\
``ZR'' state                                 \\
$t$               &0.540            &0.135   \\
$t'$              &0.305            &0.010   \\
$t''$             &0.115            &0.075   \\
\\
$d^{10}$ state                               \\
$t$               &0.290            &0.115   \\
$t'$              &0.130            &0.130   \\
$t''$             &0.045            &0.015   \\
\end{tabular}
\end{ruledtabular}
\end{table}

Knowledge of the QP dispersion enables us to study the evolution of the
{\it renormalized} FS as a function of doping.
Assuming a rigid band shift with doping, an assumption supported by independent
experiments \cite{CuO_ChemPotShift_arpes,CuO_ChemPotShift_XPS}, we plot the evolution
of the FS for {\it both} hole and electron doped cuprates in Fig.~4.
We simulate doping effects by a progressive downward shift (hole doping) and 
upward shift (electron doping) of the Fermi energy, $E_{F}$.
Once again, our results show a remarkable agreement with the experiment:
small hole pockets centered around the nodal (N) region [\,${\bf k}_{n}\!=\!(\pi/2a,\pi/2a)$\,]
comprise the ``FS'' in the deeply UD regime \cite{CuO_HP_doiron07,CuO_HP_yelland07}.
Additionally, for slightly higher $\delta$,
smaller, {\it electron}-like pockets centred around the corners of the 
Brillouin zone are also clearly resolved, see Fig.~4(a).
With further doping, these progressively evolve into a large hole-like FS
\cite{CuO_damascelli_rev}, implying a FS reconstruction close to $\delta_{opt}$. 
At a critical value $\delta\!=\!\delta_{c}$, the FS changes from hole-like ($\delta\!<\!\delta_{c}$)
to electron-like ($\delta\!>\!\delta_{c}$), in complete accord with results from ARPES 
\cite{CuO_damascelli_rev}.
Excellent agreement of the FS vis-a-vis experiment \cite{CuO_nSC_armitage02}
is also obtained for the $n$-type cuprates:
small pockets centered around ${\bf k}_{an}$ evolve into a hole-like FS with
progressive electron doping.
To our knowledge, ours are the first {\it ab initio} results capturing such effects:
hitherto, these have been (partially) described within {\it effective}, one-
\cite{kakehashi_06,CuO_FS_alex_06,dmft_rev_GK06}
or three-band \cite{CuO_pd_hanke92,CuO_pd_unger93} models with parametrized
couplings.

%% FIGURE 3
\begin{figure}[b]
\includegraphics*[angle=0,width=0.90\columnwidth]{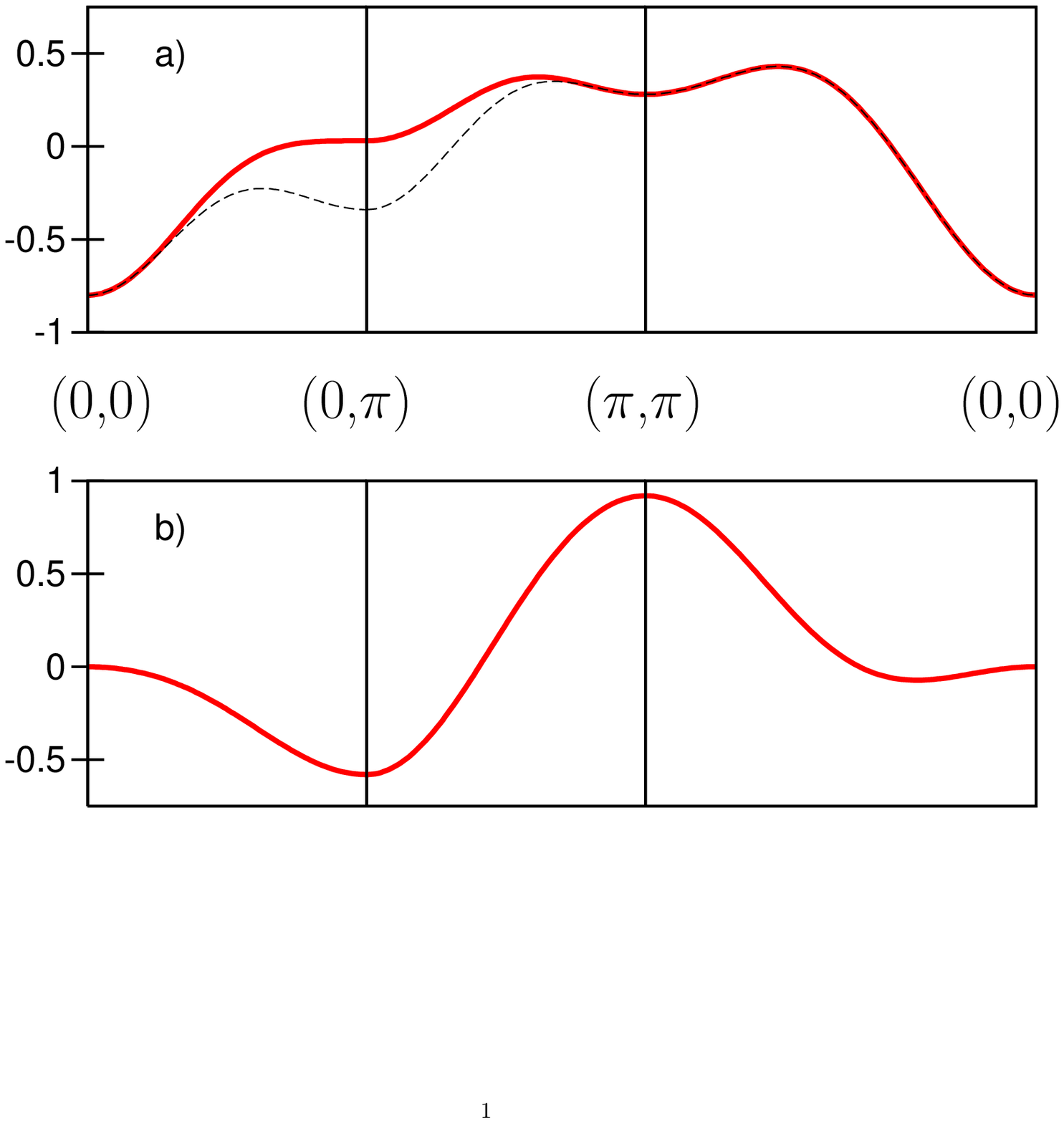}
\caption{(a): The ZR-like electron-removal band for La$_2$CuO$_4$ in the 2D
Brillouin zone, without including the interaction with the $d_{3z^2-r^2}$ hole state
(dashed line) and after including this interaction (thick red line).
For clarity, the $d_{3z^2-r^2}$ band is not shown in the figure.
The zero of energy is the value of the onsite Hamiltonian ME of the ZR state,
$H_{ii}^{\mathrm{ZR}}$.
(b): QP dispersion for the electron-addition Cu $d^{10}$ state in SrCuO$_2$.
The reference energy is the value of the onsite Hamiltonian ME,
$H_{ii}^{d^{10}}$.
Units of eV are used in both panels.
}
\end{figure}

\section{Broader Implications}

%% FIGURE 4
\begin{figure*}[!t]
\includegraphics*[width=15cm]{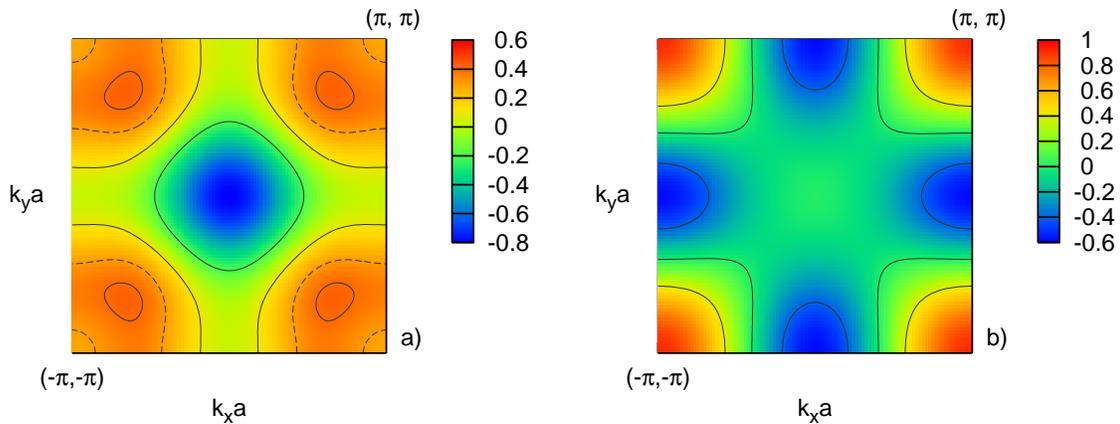}
\caption{(a): 2D colour map for the ZR-like QP dispersion in La$_2$CuO$_4$,
including the effect of the ZR--$d_{z^2}$ hole interaction.
In a rigid-band picture, the constant-energy (CE) contours (black curves) illustrate
the evolution of the FS with {\it hole doping}, from small (hole) pockets in the nodal
region for deeply UD samples  
to a large, hole-like FS at intermediate dopings and an electron-like shape in the OD
regime \cite{CuO_damascelli_rev}.
For a narrow doping interval in the UD regime, the ``FS'' is defined by eight contours
(see the dashed curves), with {\it both} hole-like sheets and electron pockets centred
at the corners of the Brillouin zone.
The latter are related to the small dip in the dispersion at the ($\pi$,$\pi$) point,
see Fig.~3(a), and have been also inferred from Hall-effect measurements
\cite{CuO_HP_leboeuf07}.
(b): Colour map for the electron-addition Cu $d^{10}$ states in SrCuO$_2$.
The CE contours reproduce the experimentally observed evolution of the FS with
{\it electron doping}, from small electron pockets in the antinodal region for the
lightly doped regime to a large, hole-like FS at high doping
\cite{CuO_damascelli_rev,CuO_nSC_armitage02}.
The same energy scales as in Fig.~3 are used.
}
\end{figure*}

Our results constitute an {\it ab initio} derivation of the ``hot-spot-cold-spot''
phenomenology \cite{CuO_hotspot_97}, also seen in cluster-DMFT work on the 2D Hubbard
model \cite{dmft_rev_GK06}.
There, the QP scattering rate is strongly ${\bf k}$-dependent as the FS is traversed.
This is mainfest in our computed FS: the pronounced dispersion of the N-QP's implies
weaker QP scattering around ${\bf k}_{n}$, in marked contrast to the antinodal (AN) region,
where strong band flattening is indicative of strong QP scattering and very short QP
lifetimes.  
Further, the renormalized $t$, $t'$, $t''$ imply {\it intrinsically} frustrated
hopping: interestingly, the importance of frustrated kinetic energy to the
high-$T_{c}$ problem has been discussed at length in the resonating valence bond
(RVB) model of Anderson \cite{CuO_RVB_97}.  
We show that {\it both} these seemingly disparate features arise 
from the same underlying microscopic mechanism:
strongly anisotropic renormalisation of carrier motion by strong, short-range
spin correlations.

Interestingly, in a Hubbard-type model, large $t'$ (or $t''$) open the door to additional
exotic phases, like $d$-wave nematic \cite{dwN_metzner_07}, $d$-density wave
\cite{dDW_chak_07} and valence-bond \cite{VBO_sachdev_01} ordered phases.
These have been invoked as possible competitors of $d$-SC in various semiphenomenological
contexts. 
Our work establishes the intimate connection between these putative instabilities
and short-range 
spin correlations characteristic of a (lightly doped) MI.

Our findings have remarkable implications for the interpretation of a host of
experiments probing the unusual ``normal''-state physical responses of the cuprates 
as a function of doping.
First, the UD $p$-type cuprates are revealed to be N-metals for small $\delta$, as
inferred by ARPES \cite{CuO_tanaka06,CuO_kanigel06}.
All the ``nodal metal'' phenomenology \cite{CuO_PALee_06} can now be provided an
{\it ab initio} justification in light of our results.
Coherent nodal QP's also naturally explain the ``good metal'' thermal conductivity in
the UD region \cite{CuO_NM_sutherland05}.
Additionally, the SdH quantum oscillation frequencies $\Omega$ \cite{CuO_HP_doiron07,CuO_HP_yelland07}
now correspond to coherent carrier orbits in the ``pockets''.
The Onsager-Lifshitz formula, $\Omega=\Phi_{0}A/(2\pi)^{2}$, with $\Phi_{0}=hc/e$
the flux quantum and $A=2\pi^{2}x/4a^{2}$ the area of the pocket for $x$ carriers
in ${\bf k}$-space, tells us that $\Omega(x)\!\sim\!x$.
Our finding of additional {\it electron}-like pockets around $(\pi,\pi)$ in the hole doped 
case offers a route toward a resolution of one of the central controversies surrounding recent 
SdH experiments, where $x_{{\rm SdH}}\!=\!0.15$ for the $\delta\!=\!0.10$ cuprates
\cite{CuO_HP_doiron07}.  
While this is irreconcilable with theories having only hole pockets, our finding of
additional electron-like sheets can reconcile the SdH results with the Luttinger sum
rule for $\delta\!\approx\!0.1$ \cite{CuO_HP_doiron07,CuO_HP_yelland07}.
Moreover, the Hall constant $R_{H}(x)$ is now expected to track the evolution of the 
renormalised FS with doping.
Depending upon the concentrations of hole and electron carriers $n_{h}$ and $n_{e}$, with
$n_{h}\!>\!n_{e}$, and the mobilities $\mu_{h}$ and $\mu_{e}$ \cite{CuO_HP_leboeuf07}, 
$R_H$ may change sign from hole-like to electron-like with $T$, reconciling the SdH
\cite{CuO_HP_doiron07,CuO_HP_yelland07} and Hall-effect \cite{CuO_HP_leboeuf07} data.

Finally, what about superconductivity\,?
Given the small number, $\delta$, of nodal QP's in the UD regime, $d$-SC will result from
pairing of these quasicoherent fermionic entities. 
Without going into the nature of the pairing mechanism, the above implies that 
$T_{c}(\delta)\!\sim\!\delta$ for $\delta\!\ll\!1$.
So the superfluid density at $T\!=\!0$ will decrease {\it linearly} as $\delta$ is reduced,
implying that upon underdoping $d$-SC will be progressively destroyed by order parameter 
phase fluctuations, which grow as the MI is approached. 
This implies non-BCS
(large) values of $2\Delta/k_{B}T_{c}$ and strong vortex-liquid like effects above
$T_{c}$ for UD cuprates.
These have indeed been invoked in connection with the anomalous, ``giant'' Nernst effect
in UD superconductors \cite{CuO_NE_Ong_06}.
However, once a full FS develops, $d$-SC would be expected to revert back to a more
conventional BCS-like variety \cite{CuO_LeTacon_06}, with identical scaling for the N and
AN gaps.

\section{Conclusions}

To summarize, we have implemented a first-principles, wavefunction-based 
calculation of correlated hole and electron-addition quasiparticle states in
layered cuprates.
In addition to quantitatively describing the dispersion of the ZR-like band,  
our work reproduces the FS evolution as a function of doping, in remarkable
agreement with a host of recent ARPES and quantum oscillation experiments.
Our finding of large longer-range effective hoppings implies intrinsically frustrated 
carrier kinetic energy, in agreement with Anderson's RVB ideas
\cite{CuO_RVB_97}.
The very different behavior of hole and electron doped cuprates is clearly 
manifested as originating from very different quantum chemical and spin correlation
``backgrounds''.
Seen from this perspective, the FS ``reconstruction'' with doping \cite{CuO_HP_leboeuf07},
as well as the famed nodal-antinodal dichotomy in the UD systems, are both
understood in terms of ${\bf k}$-space differentiation of QP states in the 2D,
doped MI.
Phenomenologically, the computed evolution of the FS with $\delta$ goes hand-in-hand with 
the observed evolution of $d$-SC from a strongly non-BCS, phase fluctuation dominated
type, to a more conventional BCS type with progressive doping, benchmarking the crucial
relevance of fermiology in cuprates.

{\small
\subsection{--- \ Methods \ ---}

In the CASSCF approach \cite{casscf_roos80,QC_book_00}, the wavefunction is written
as a linear combination of configuration state functions (CSF's) $\vert m\rangle$,
$|\Psi \rangle = \sum_{m} C_{m}\:\vert m \rangle$\,.
These CSF's are spin- (and symmetry-) adapted combinations of Slater determinants (SD's),
i.\,e., eigenfunctions of the operators for the projected and total spins.
In turn, the SD's are constructed from a set of real and orthonormal spin
orbitals $\{\phi _{{\sc p}}({\bf r},\sigma)\}$, where ${\bf r}$ and $\sigma$ are the spatial
and spin coordinates, respectively.
In this work, an initial guess for these orbitals is obtained from a Hartree-Fock
calculation for an hypothetical Cu $3d^{10}$, O $2p^6$ closed-shell configuration of
the Cu and O species.

In determining the CASSCF wavefunction, the orbitals are variationally optimized
simultaneously with the coefficients of the CSF's.
The orbitals employed for expressing the wavefunction are thus the optimal orbitals for
the state at hand and do not introduce a bias toward a particular configuration.
Three different sets of orbitals are used in CASSCF:
(i) the inactive levels, doubly occupied in all configurations,
(ii) the virtual orbitals, unoccupied in all configurations, and
(iii) the active orbital set, where no occupancy restrictions are imposed.
For undoped cuprates, with formal Cu $3d^{9}$ and O $2p^{6}$ valence
states, the active space would include the partially occupied in-plane Cu
$3d_{x^{2}-y^{2}}$ orbitals.
Such a CAS wavefunction is similar to the variational wavefunction used in numerical
studies of the 2D, one-band Hubbard model \cite{HubbM_sorella_05}.
The main difference is that {\it all} integrals, including inter-site Coulomb and
exchange terms, are computed here in a totally {\it ab initio} way.
In particular, the lower, completely filled levels, e.\,g., the O $2s$ and $2p$ orbitals,
do affect (i.e., screen) the actual interactions among the active electrons 
by readjusting themselves to fluctuations within the active orbital space.

If extra holes are created, the active space must be enlarged with orbitals from the
inactive group.
Each additional doped hole requires one orbital to be transferred from the inactive to
the active space.
For the lowest electron-removal state, for example, the orbital added to the active
space turns into a ZR-type $p$-$d$ composite \cite{ZR_88} in the variational
calculation, localized on a given CuO$_4$ plaquette.
With regard to the electron-addition conduction-band states, these turn out to have Cu
$3d^{10}$ character and, in a first approximation, an active orbital space including
only the $3d_{x^{2}-y^{2}}$ levels would suffice.

So-called dynamic correlation effects \cite{QC_book_00} for the onsite matrix elements
$H_{ii}$ were computed using second-order perturbation theory (the CASPT2 method
\cite{CASPT2_90}).
The Cu $3d$ and O $2s$, $2p$ electrons on five plaquettes (i.e., the ZR plaquette and
the two apical ligands for that plaquette plus the four NN plaquettes) were correlated.
All calculations were performed with the QC software {\sc molcas} \cite{molcas6}.
For the ions of the plaquettes directly involved in the hopping process, all-electron
basis sets (BS's) of triple-zeta quality were applied. 
These were Gaussian-type atomic-natural-orbital BS's from the \textsc{molcas} library
\cite{molcas6}, with the following contractions \cite{QC_book_00}:
Cu ($21s15p10d$)/[$5s4p3d$] and O ($14s9p$)/[$4s3p$].
The core electrons of the remaining ions of each cluster, see Fig.~2, were represented by
effective core potentials (ECP's), i.e., Cu ECP's plus valence double-zeta BS's  
\cite{ECPs_Cu} and O ECP's with triple-zeta BS's \cite{ECPs_O}.
To describe the finite charge distribution at the sites in the immediate neighborhood
of the cluster, we model those ions by effective ion potentials \cite{TIPs_CuSr}.
Beyond these neighbors, we use large arrays of point charges (PC's) that reproduce the
Madelung field within the cluster region.
Apical ligands are explicitly included in our calculations only for the ``active''
plaquettes.
Other apex oxygens are represented by formal PC's.
That the charge populations of the Cu $3d$ and active O $2p$ orbitals are not sensitive
to the size and shape of the clusters we use was shown in ref.~\cite{CuO_hozoi_07}.
We employed the crystal structure measured by Cava {\it et al.}~\cite{LaCuO_xrd_cava87}
for La$_2$CuO$_4$ and by Smith {\it et al.}~\cite{SrCuO_JBG_91} for SrCuO$_2$.
}


\begin{thebibliography}{99}

\bibitem{CuO_PALee_06} Lee, P. A., Nagaosa, N. \& Wen, X.-G.,
Doping a Mott insulator: physics of high-temperature superconductivity.
{\it Rev. Mod. Phys.} {\bf 78}, 17 (2006).

\bibitem{CuO_NE_Ong_06} Wang, Y., Li., L \& Ong, N. P.,
Nernst effect in high-$T_c$ superconductors.
{\it Phys. Rev. B} {\bf 73}, 024510 (2006).

\bibitem{CuO_PG_varma06} Varma, C. M.,
Theory of the pseudogap state of the cuprates.
{\it Phys. Rev. B} {\bf 73}, 155113 (2006).

\bibitem{CuO_QC_vdMarel_03} van der Marel, D. {\it et al.}, 
Quantum critical behaviour in a high-$T_c$ superconductor.
{\it Nature} {\bf 425}, 271-274 (2003). 

\bibitem{CuO_damascelli_rev} Damascelli, A., Hussain, Z. \& Shen, Z.-X.,
Angle-resolved photoemission studies of the cuprate superconductors.
{\it Rev. Mod. Phys.} {\bf 75}, 473-541 (2003).

\bibitem{CuO_tanaka06} Tanaka, K. {\it et al.},
Distinct Fermi-momentum-dependent energy gaps in deeply underdoped Bi2212.
{\it Science} {\bf 314}, 1910-1913 (2006).

\bibitem{CuO_kanigel06} Kanigel, A. {\it et al.},
Evolution of the pseudogap from Fermi arcs to the nodal liquid.
{\it Nature Physics} {\bf 2}, 447-451 (2006).

\bibitem{CuO_HP_doiron07} Doiron-Leyraud, N. {\it et al.},
Quantum oscillations and the Fermi surface in an underdoped high-$T_c$ superconductor.
{\it Nature} {\bf 447}, 565-568 (2007).

\bibitem{CuO_HP_yelland07} Yelland, E. A. {\it et al.},
Quantum Oscillations in the Underdoped Cuprate YBa$_2$Cu$_4$O$_8$.
arXiv:0707.0057v1 (unpublished).

\bibitem{CuO_OD_hussey06} Abdel-Jawad, M. {\it et al.},
Anisotropic scattering and anomalous normal-state transport in a high-temperature
superconductor.
{\it Nature Physics} {\bf 2}, 821-825 (2006).

\bibitem{CuO_HP_leboeuf07}  LeBoeuf, D. {\it et al.},
Electron pockets in the Fermi surface of hole-doped high-$T_c$ superconductors.
{\it Nature} {\bf 450}, 533-536 (2007).

\bibitem{CuO_RVB_97} Anderson, P. W., \textit{The Theory of Superconductivity in the
High-$T_{c}$ Cuprates} (Princeton Univ. Press, 1997).

\bibitem{FMpolaron_emery88} Emery, V. J. \& Reiter, G.,
Mechanism for high-temperature superconductivity.
{\it Phys. Rev. B} {\bf 38}, 4547-4556 (1988).

\bibitem{ZR_88} Zhang, F. C. \&  Rice, T. M.,
Effective Hamiltonian for the superconducting Cu oxides.
{\it Phys. Rev. B} {\bf 37}, 3759-3761 (1988).

\bibitem{CuO_hozoi_07} Hozoi, L., Nishimoto, S. \& de Graaf, C.,
Renormalization of quasiparticle hopping integrals by spin interactions in layered
copper oxides.
{\it Phys. Rev. B} {\bf 75}, 174505 (2007).

\bibitem{dmft_rev_GK06} Kotliar, G. {\it et al.},
Electronic structure calculations with dynamical mean-field theory.
{\it Rev. Mod. Phys.} {\bf 78}, 865 (2006).

\bibitem{dmft_U_imada05} Solovyev, I. V. \& Imada, M.,
{\it Phys. Rev. B} {\bf 71}, 045103 (2005).

\bibitem{fulde_wf_02} Fulde, P.,
Wavefunction methods in electronic-structure theory of solids.
{\it Adv. Phys.} {\bf 51}, 909-948 (2002).

\bibitem{QC_book_00} Helgaker, T., J\o rgensen, P. \& Olsen, J.,
\textit{Molecular Electronic-Structure Theory}
(Wiley, Chichester, 2000).

\bibitem{casscf_roos80} Roos, B. O., Taylor, P. R. \& Siegbahn, P. E. M.,
A complete active space SCF method (CASSCF) using a density matrix formulated
super-CI approach.
{\it Chem. Phys.} {\bf 48}, 157-173 (1980).

\bibitem{thesis_Coen} de Graaf, C.,
Local excitations and magnetism in late transition metal oxides,
Ph. D. Thesis, University of Groningen, 1998
(available at \url{http://theochem.chem.rug.nl/publications/PDF/ft325.pdf}).

\bibitem{DE_alex06} Stoyanova, A., Sousa, C., de Graaf, C. \& Broer, R.,
Hopping matrix elements from first-principles studies of overlapping fragments:
double exchange parameters in manganites.
{\it Int. J. Quantum Chem.} {\bf 106}, 2444-2457 (2006).

\bibitem{HubbM_sorella_05} Capello, M. {\it et al.}, 
Variational description of Mott insulators.
{\it Phys. Rev. Lett.} {\bf 94}, 026406 (2005).

\bibitem{CuO_oka_01} Pavarini, E. {\it et al.},
Band-structure trend in hole-doped cuprates and correlation with $T_{c\ max}$.
{\it Phys. Rev. Lett.} {\bf 87}, 047003 (2001).

\bibitem{CBs_uwe_06} Birkenheuer, U., Fulde, P. \&  Stoll, H.,
A simplified method for the computation of correlation effects on the band structure
of semiconductors.
{\it Theor. Chem. Acc.} {\bf 116}, 398-403 (2006).

\bibitem{CBs_MgO_07} Hozoi, L. {\it et al.},
Ab initio wavefunction-based methods for excited states in solids:
correlation corrections to the band structure of ionic oxides.
{\it Phys. Rev. B} {\bf 76}, 085109 (2007).

\bibitem{SI_Malmqvist_86} Malmqvist, P.-\AA .,
Calculation of transition density matrices by nonunitary orbital transformations.
{\it Int. J. Quantum Chem.} {\bf 30}, 479 (1986).

\bibitem{CASPT2_90} Andersson, K. {\it et al.},
Second-order perturbation theory with a CASSCF reference function.
{\it Phys. Chem.} {\bf 94}, 5483-5488 (1990).

\bibitem{SrCuO_JBG_91} Smith, M. G. {\it et al.},
Electron-doped superconductivity at 40 K in the infinite-layer compound
Sr$_{1-y}$Nd$_y$CuO$_2$.
{\it Nature} {\bf 351}, 549-551 (1991).

\bibitem{CuO_LH_MSL_07} Hozoi, L. \& Laad, M. S.,
Quasiparticle bands in cuprates by quantum-chemical methods:
towards an ab initio description of strong electron correlations.
{\it Phys. Rev. Lett.} {\bf 99} 256404 (2007).

\bibitem{CuO_ARPES_ino00} Ino, A. {\it et al.},
Electronic structure of La$_{2-x}$Sr$_x$CuO$_4$ in the vicinity of the
superconductor-insulator transition.
{\it Phys. Rev. B} {\bf 62}, 4137-4141 (2000).

\bibitem{CuO_WF_fink07} Inosov, D. S. {\it et al.},
Momentum and energy dependence of the anomalous high-energy dispersion in the
electronic structure of high temperature superconductors.
{\it Phys. Rev. Lett.} {\bf 99}, 237002 (2007).

\bibitem{kakehashi_06} Kakehashi, Y. \& Fulde, P.,
Nonlocal excitation spectra in the two-dimensional doped Hubbard model.
{\it J. Phys. Soc. Jpn.} {\bf 76}, 074702 (2007).

\bibitem{CuO_ChemPotShift_arpes} Ronning, F. {\it et al.},
Evolution of a metal to insulator transition in Ca$_{2-x}$Na$_x$CuO$_2$Cl$_2$ as 
seen by angle-resolved photoemission.
{\it Phys. Rev. B} {\bf 67}, 165101 (2003).

\bibitem{CuO_ChemPotShift_XPS}
Hashimoto, M. {\it et al.},
Doping evolution of the electronic structure in the single-layer cuprates
Bi$_{2}$Sr$_{2-x}$La$_x$CuO$_{6+\delta}$: comparison with other single-layer 
cuprates.
arXiv:0801.0782 (unpublished).
%% Yagi, H. {\it et al.},
%% Chemical potential shift in lightly doped to optimally doped
%% Ca$_{2-x}$Na$_x$CuO$_2$Cl$_2$.
%% {\it Phys. Rev. B} {\bf 73}, 172503 (2006).


\bibitem{CuO_nSC_armitage02} Armitage, N. P. {\it et al.},
Doping dependence of an $n$-type cuprate superconductor investigated by 
angle-resolved photoemission spectroscopy.
{\it Phys. Rev. Lett.} {\bf 88}, 257001 (2002).

\bibitem{CuO_FS_alex_06} Macridin, A. {\it et al.},
Pseudogap and antiferromagnetic correlations in the Hubbard model.
{\it Phys. Rev. Lett.} {\bf 97}, 036401 (2006).

\bibitem{CuO_pd_hanke92} Dopf, G. {\it et al.},
Direct comparison of angle-resolved photoemission and numerical simulations for
high-$T_c$ superconductors. 
{\it Phys. Rev. Lett.} {\bf 68}, 2082-2085 (1992).

\bibitem{CuO_pd_unger93} Unger, P. \& Fulde, P.,
Spectral function of holes in the Emery model. 
{\it Phys. Rev. B} {\bf 48}, 16607-16622 (1993).

\bibitem{CuO_hotspot_97} Stojkovi\'{c}, B. P. \& Pines, D.,
Theory of the longitudinal and Hall conductivities of the cuprate superconductors. 
{\it Phys. Rev. B} {\bf 55}, 8576-8595 (1997). 

\bibitem{dwN_metzner_07} Yamase, H. \& Metzner, W.,
Competition of Fermi surface symmetry breaking and superconductivity.
{\it Phys. Rev. B} {\bf 75}, 155117 (2007).

\bibitem{dDW_chak_07} Chakravarty, S. \& Kee, H.-Y.,
Fermi pockets and quantum oscillations of the Hall coefficient in high temperature
superconductors.
arXiv:0710.0608 (unpublished).

\bibitem{VBO_sachdev_01} Park, K. \& Sachdev, S.,
Bond-operator theory of doped antiferromagnets: from Mott insulators with
bond-centered charge order to superconductors with nodal fermions. 
{\it Phys. Rev. B} {\bf 64}, 184510 (2001).
 
\bibitem{CuO_NM_sutherland05} Sutherland, M. {\it et al.},
Delocalized fermions in underdoped cuprate superconductors.
{\it Phys. Rev. Lett.} {\bf 94}, 147004 (2005).

\bibitem{CuO_LeTacon_06} Le Tacon, M. {\it et al.},
Two energy scales and two distinct quasiparticle dynamics in the superconducting
state of underdoped cuprates.
{\it Nature Physics} {\bf 2}, 537-543 (2006).

\bibitem{molcas6} \textsc{molcas 6}, Department of Theoretical
Chemistry, University of Lund, Sweden.

\bibitem{ECPs_Cu} Seijo, L., Barandi\'{a}ran, Z. \& Huzinaga, S.,
The ab initio model potential method. First series transition metal elements.
{\it J. Chem. Phys.} {\bf 91}, 7011-7017 (1989).

\bibitem{ECPs_O} Bergner, A. {\it et al.},
Ab initio energy-adjusted pseudopotentials for elements of groups 13-17.
{\it Mol. Phys.} {\bf 80}, 1431-1441 (1993).

\bibitem{TIPs_CuSr} Illas, F., Rubio, J. \& Barthelat, J. C.,
Cu as a one-electron atom: molecular structure and dissociation energy of CuOH.
{\it Chem. Phys. Lett.} {\bf 119}, 397-402 (1985).

\bibitem{LaCuO_xrd_cava87} Cava, R. J., Santoro, A., Johnson, D. W. \&
Rhodes, W. W.,
Crystal structure of the high-temperature superconductor
La$_{1.85}$Sr$_{0.15}$CuO$_4$ above and below $T_c$. 
{\it Phys. Rev. B} {\bf 35}, 6716-6720 (1987).

\end{thebibliography}
\end{document}